\documentclass[sigconf]{acmart}

\AtBeginDocument{%
  \providecommand\BibTeX{{%
    \normalfont B\kern-0.5em{\scshape i\kern-0.25em b}\kern-0.8em\TeX}}}

\copyrightyear{2021}
\acmYear{2021}
\setcopyright{acmlicensed}\acmConference[KDD '21]{Proceedings of the 27th ACM SIGKDD Conference on Knowledge Discovery and Data Mining}{August 14--18, 2021}{Virtual Event, Singapore}
\acmBooktitle{Proceedings of the 27th ACM SIGKDD Conference on Knowledge Discovery and Data Mining (KDD '21), August 14--18, 2021, Virtual Event, Singapore}
\acmPrice{15.00}
\acmDOI{10.1145/3447548.3467089}
\acmISBN{978-1-4503-8332-5/21/08}

\usepackage{subcaption}
\usepackage{enumitem}
\usepackage{bm}
\usepackage{url}

\makeatletter
\def\blfootnote{\gdef\@thefnmark{}\@footnotetext}
\makeatother

\DeclareMathOperator*{\argmax}{arg\,max~}
\DeclareMathOperator*{\argmin}{arg\,min~}

\newcommand{\data}{\mathcal{D}}

\newcommand{\E}{\mathbb{E}}
\newcommand{\Em}{\mathop{{}\mathbb{E}}}

\newcommand{\N}{\mathcal{N}}
\newcommand{\KL}{\mathrm{KL}}

\newcommand{\slf}{h}
\newcommand{\defeq}{\mathrel{\overset{\makebox[0pt]{\mbox{\normalfont\tiny\sffamily def}}}{=}}}
\newcommand{\eqn}[1]{Eq.~(\ref{eqn:#1})}
\newcommand{\eqnt}[2]{Eq.~(\ref{eqn:#1} \& \ref{eqn:#2})}
\newcommand{\secref}[1]{Sec.~\ref{sec:#1}}
\newcommand{\figref}[1]{Fig.~\ref{fig:#1}}
\newcommand{\figreft}[2]{Fig.~\ref{fig:#1} \& \ref{fig:#2}}
\newcommand{\tabref}[1]{Table~\ref{table:#1}}
\newcommand{\blue}[1]{\textcolor{blue}{#1}}
\newcommand{\eyx}{\E[y|x]}

\graphicspath{{./figures/}}

\begin{document}
\fancyhead{}

\title[Deep Uncertainty-Aware Learning]{Exploration in Online Advertising Systems with\\ Deep Uncertainty-Aware Learning}

\author{Chao Du, Zhifeng Gao, Shuo Yuan, Lining Gao, Ziyan Li, Yifan Zeng, Xiaoqiang Zhu\\ Jian Xu, Kun Gai, Kuang-chih Lee}
\email{{cangyun.dc, zhifeng.gzf, yuanshuo.ys, lining.gln, liza.lzy, xiaoqiang.zxq, xiyu.xj, kuang-chih.lee}@alibaba-inc.com}
\affiliation{%
\institution{Alibaba Group}
\city{Beijing}
\country{China}
}
\renewcommand{\shortauthors}{Du et al.}

\begin{abstract}
  Modern online advertising systems inevitably rely on personalization methods, such as click-through rate (CTR) prediction. Recent progress in CTR prediction enjoys the rich representation capabilities of deep learning and achieves great success in large-scale industrial applications. However, these methods can suffer from lack of \emph{exploration}. Another line of prior work addresses the \emph{exploration-exploitation} trade-off problem with contextual bandit methods, which are recently less studied in the industry due to the difficulty in extending their flexibility with deep models. In this paper, we propose a novel \emph{Deep Uncertainty-Aware Learning} (DUAL) method to learn CTR models based on Gaussian processes, which can provide predictive \emph{uncertainty estimations} while maintaining the flexibility of deep neural networks. DUAL can be easily implemented on existing models and deployed in real-time systems with minimal extra computational overhead. By linking the predictive uncertainty estimation ability of DUAL to well-known bandit algorithms, we further present \emph{DUAL-based Ad-ranking strategies} to boost up long-term utilities such as the social welfare in advertising systems. Experimental results on several public datasets demonstrate the effectiveness of our methods. Remarkably, an online A/B test deployed in the Alibaba display advertising platform shows an $8.2\%$ social welfare improvement and an $8.0\%$ revenue lift.
  \blfootnote{Correspondence to: C. Du <duchao0726@gmail.com; cangyun.dc@alibaba-inc.com>}
\end{abstract}




\maketitle

\begin{figure}[t!]
  \centering
  \includegraphics[width=\linewidth, height=.23\linewidth]{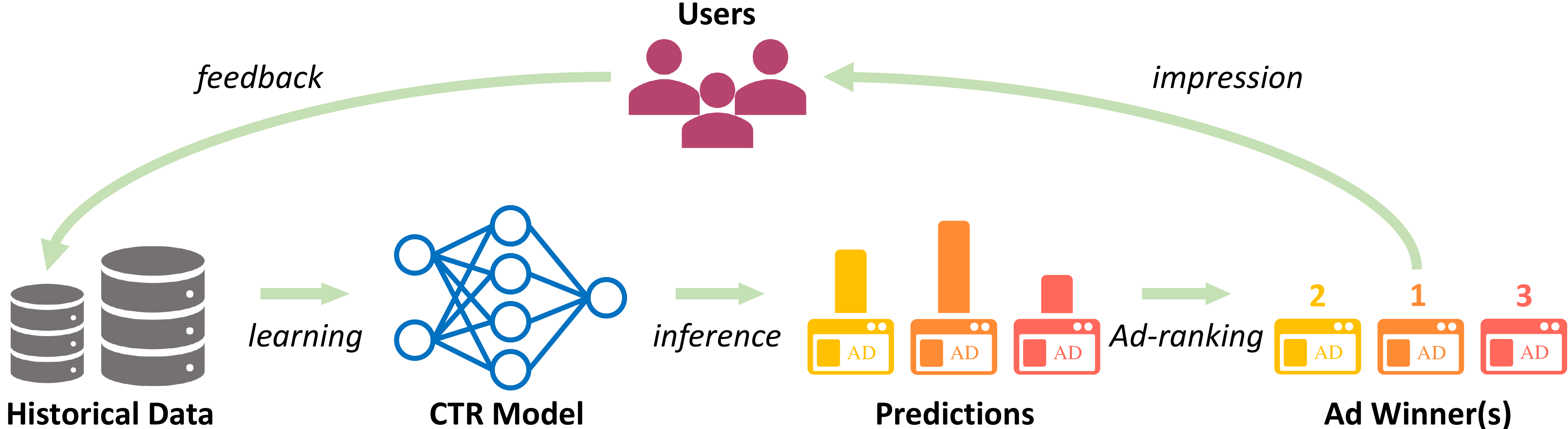}
  \vspace{-0.6cm}
  \caption{An illustration of CTR prediction models and Ad-ranking strategies in common online advertising systems.
   }
  \label{fig:loop}
  \Description{}
  \vspace{-0.4cm}
\end{figure}

\section{Introduction}
\label{sec:intro}

Online advertising~\cite{broder2008computational} has become one of the most important business in today's e-commerce platforms, social networks, and news/movie content providers, where billions of Internet users can be reached every day.
In current online advertising markets, cost-per-click (CPC) campaigns are widely preferred, where advertisers pay for each click on their advertisements (Ads).
The effectiveness of advertising systems therefore largely depends on personalization techniques,
such as click-through rate (CTR) prediction,
which captures user interests and helps distribute online traffic efficiently,
and potentially reduces unpleasant user experiences.

Recently, deep learning-based CTR prediction methods have achieved great success~\cite{cheng2016wide,qu2016product,guo2017deepfm,zhou2018deep,zhou2019deep} due to their ability to learn rich representations of user behaviors and Ad characteristics.
Such models are usually trained in a purely supervised learning manner, where the training data (i.e., Ad impressions and user feedback) are sequentially obtained from online serving.
Based on the CTR predictions,
a common \emph{Ad-ranking strategy} derived from the widely used generalized second-price auction~\cite{edelman2007internet} is to display Ads according to the ranking of predicted eCPMs (effective cost per mille),
which in turn produces the data for further updates of the CTR model (as illustrated in \figref{loop}).
However, due to lack of \emph{exploration}~\cite{chaney2018algorithmic} (as detailed in \secref{back}), these models can gradually focus on a small subset of Ad candidates
and miss out on opportunities to find the unknown better ones, potentially leading to suboptimal predictive performance and unsatisfying social welfare~\cite{edelman2007internet}.

Exploring in advertising systems can be very expensive.
In contrast to supervised learning, another line of prior work in recommendation systems (which is closely related to advertising systems) is formulated as a \emph{contextual bandit} problem~\cite{sutton2018reinforcement,li2010contextual},
in which the system sequentially displays items to the users based on contextual information of the user and items, and updates its item-selection strategy while explicitly balancing the \emph{exploration-exploitation} trade-off~\cite{balabanovic1998exploring}.
Such methods have proven beneficial for long-term utilities theoretically and practically~\cite{chu2011contextual,li2010contextual}.
However, 
most of the existing works~\cite{li2010contextual,li2011unbiased,dumitrascu2018pg} are limited by flexibility due to their linear or logistic assumptions of the true CTR functions,
making them unable to match the advantages of deep CTR models
and therefore limited in today's real-world systems.

To conjoin the best of both worlds,
we propose a novel \emph{Deep Uncertainty-Aware Learning} (DUAL) approach to CTR prediction, which can be adapted to contextual bandit methods
while maintaining the powerful representation capabilities of deep neural networks.
The core idea is to learn deep CTR models that can provide \emph{uncertainty estimations} of the predictions, which is the key to navigating the \emph{exploration} in contextual bandit problems.
In DUAL, 
possessing a Bayesian perspective of CTR modeling,
we impose a Gaussian process (GP)~\cite{rasmussen2003gaussian} prior distribution to the unknown true CTR function and obtain the predictions as well as the uncertainty estimations from the posterior given the observed data.
We present practical learning and inference algorithms for DUAL and an efficient implementation so that it can be easily deployed in real-time systems with minimal extra computational overhead.
By combining the uncertainty estimations from DUAL with widely-known bandit algorithms such as UCB~\cite{auer2002using} and Thompson Sampling~\cite{chapelle2011empirical}, we obtain new \emph{DUAL-based Ad-ranking strategies},
which manage to balance the exploration-exploitation trade-off and improve long-term utilities.
Finally, we demonstrate the effectiveness of our proposed methods in terms of predictive performance and long-term social welfare on several public datasets, and identify their good practices in the Alibaba display advertising platform.

To sum up, our main contributions are as follows:
\begin{itemize}
  \item We present a novel approach to learning deep CTR prediction models, named \emph{Deep Uncertainty-Aware Learning} (DUAL), which can provide efficient uncertainty estimations along with the predictions and is compatible with existing deep CTR models favored in large-scale industrial applications.
  \item We propose new \emph{DUAL-based Ad-ranking strategies} by combining CTR models learned by DUAL with bandit algorithms, to efficiently balance the exploration-exploitation trade-off and optimize long-term utilities in advertising systems.
  \item We conduct extensive experiments on several public datasets. Results demonstrate the effectiveness of the proposed DUAL approach for learning deep CTR models and DUAL-based strategies for long-term social welfare optimization. 
  \item Remarkably, a two-week online A/B test deployed in the Alibaba display advertising platform shows an $8.2\%$ social welfare improvement and an $8.0\%$ revenue lift.
\end{itemize}




\vspace{-.2cm}
\section{Background and Motivation}
\label{sec:back}

We review the importance of CTR prediction in advertising systems and the common manner for learning CTR models. Then we discuss the assumptions and drawbacks behind, which motivate us to move towards a contextual bandit formulation of online advertising.


\vspace{-.2cm}
\subsection{Online Advertising Systems}
\label{sec:back:adsys}

Currently, most cost-per-click (CPC) advertising platforms sell advertising spaces through online auctions, such as the generalized second-price (GSP) auction~\cite{edelman2007internet}.
Under GSP auctions, Ads are displayed according to the ranking of their expected values of impression, each of which is referred to as the eCPM\footnote{The precise definition of eCPM is the effective cost per thousand impressions. Here we have slightly abused the expected value of a single impression and the eCPM.} and defined as the product of the advertiser's bid price and the CTR.
Such a mechanism is shown to have an equilibrium maximizing the \emph{social welfare}~\cite{edelman2007internet}, i.e., the total value obtained by all advertisers,
and thus promotes a long-term win-win situation for advertisers and platforms.
However, since the true CTRs are inaccessible, the eCPMs are usually estimated with predicted CTRs.
Therefore, CTR prediction models play an important role in today's real-world advertising systems.

\vspace{-.2cm}
\subsection{CTR Prediction}
\label{sec:back:ctr}

CTR prediction models are usually learned from data collected during online serving.
The contextual information of each Ad impression and the user feedback are recorded as a data sample.
Specifically, this information includes \emph{User Info}, which contains the user's profile and historical behaviors; \emph{Ad Info}, which characterizes the content of the Ad; and \emph{Env Info}, which records the environmental information such as the scene and the timestamp of the impression.

For algorithmic modeling, the contextual information of each data sample is usually represented as a feature vector. 
To facilitate subsequent analysis, we introduce \emph{user feature} denoted by $x_{_U}$, which is the vectorized representation of the \emph{User Info},
and let $\mathcal{X}_{_U}$ denote the space of user features, i.e., $x_{_U}\in\mathcal{X}_{_U}$.
The \emph{Ad feature} $x_{_A}$, \emph{Env feature} $x_{_E}$, and their corresponding feature spaces $\mathcal{X}_{_A}$ and $\mathcal{X}_{_E}$ are defined by analogy.
We denote the feature vector of the data sample by the concatenation $x=[x_{_U}^\top,x_{_A}^\top,x_{_E}^\top]^\top$.
We assume the joint feature space $\mathcal{X}=\mathcal{X}_{_U}\times\mathcal{X}_{_A}\times\mathcal{X}_{_E}$ has a dimension\footnote{
In large-scale advertising systems, feature vectors are usually extremely high-dimensional (e.g., $d\sim 10^7$) and sparse due to the huge amount of discrete IDs of users and Ads.
A common practice is to translate them into lower-dimensional dense vectors through embeddings.
We refer to \citet{zhou2018deep} for a detailed description.
}
of $d$, i.e., $\mathcal{X}\subset\mathbb{R}^d$.
The feedback is regarded as the binary label $y\in\mathcal{Y}=\{0,1\}$, whose value takes $1$ for positive feedback (i.e., click) and $0$ otherwise.
Finally, the dataset $\data=\{(x_i,y_i)\}_{i=1}^{N}$ for training CTR models is composed of all the observed data samples.



Most existing works~\cite{zhou2018deep,cheng2016wide} tackle the CTR prediction task in a purely supervised learning manner, i.e., learning a function $\slf$ mapping from 
$\mathcal{X}$ to the CTR space $[0,1]$ (or the logit space $\mathbb{R}$) by minimizing the empirical risk.
Formally, let $\mathcal{H}\subset\mathbb{R}^{\mathcal{X}}$ be the hypothesis space\footnote{$\mathbb{R}^{\mathcal{X}}$ denotes the set of all functions $f:\mathcal{X}\rightarrow\mathbb{R}$.} and $\ell:(\mathcal{X}\times\mathcal{Y})\times\mathcal{H}\rightarrow\mathbb{R}$ be the loss function. Given the training data $\data$, the learning problem is to find an empirical risk minimizer (ERM) $\hat{\slf}\in\mathcal{H}$ that minimizes the empirical risk $\hat{L}(\slf)$:\\[-.3cm]
\begin{equation}\label{eqn:ErRM}
  \hat{L}(\slf) \defeq {} \frac{1}{N}\sum{}_{i=1}^N\ell((x_i,y_i),\slf),\quad
  \hat{\slf} \in \argmin_{\slf\in\mathcal{H}} \hat{L}(\slf).
\end{equation}
For CTR prediction, the loss function is usually defined as the negative log-likelihood of the sample $(x,y)$ induced by the model $\slf$, i.e., $\ell((x,y),\slf)=-\log p(y|x;\slf)$ and\\[-.3cm]
\begin{equation}
  p(y|x;\slf) = y\cdot \sigma(\slf(x)) + (1-y)\cdot(1-\sigma(\slf(x))),
\end{equation}
where we have assumed without loss of generality that the model $\slf:\mathcal{X}\rightarrow\mathbb{R}$ outputs a logit value and the predicted CTR can be obtained by applying the sigmoid transformation $\sigma(\cdot)=\frac{1}{1+e^{-(\cdot)}}$ to the logit.
Finally,
practitioners adopt large-scale optimization algorithms, such as stochastic gradient descent (SGD)~\cite{bottou2010large}, to obtain an (approximate) ERM $\hat{\slf}$ and make further use, e.g., online serving.

\vspace{-.2cm}
\subsection{Motivation}
\label{sec:back:drawback}
We discuss several assumptions in advertising systems built upon CTR prediction models.
First, the \emph{Ad-ranking strategy} derived from the GSP auction assumes perfect eCPM estimations.
In other words, the system implements GSP auctions only when the CTR prediction model $\hat{h}$ is well-behaved.
Second, the effectiveness of the CTR model $\hat{\slf}$ during online serving (which can be view as the test phase) relies on a key assumption behind supervised learning:
the training and test samples are drawn i.i.d. from some fixed underlying data-generating distribution $p_{_{\mathcal{X}\times\mathcal{Y}}}$ over $\mathcal{X}\times\mathcal{Y}$.
Under such an assumption,
the best possible model we can aspire to is the expected risk minimizer $\slf^*$, which minimizes the expected risk $L(\slf)$:\\[-.3cm]
\begin{equation}\label{eqn:EpRM}
  L(\slf) \defeq \E_{(x,y)\sim p_{_{\mathcal{X}\times\mathcal{Y}}}}\ell((x,y),\slf),\quad \slf^* \in \argmin_{\slf\in\mathcal{H}} L(\slf),
\end{equation}
and the uniform convergence bounds~\cite{vapnik2013nature} guarantee that the performance gap
$L(\hat{\slf})-L(\slf^*)$ will not be too large with high probability, which means the learned model $\hat{\slf}$ is reliable.

\begin{figure}[t]
  \centering
  \includegraphics[width=\linewidth,height=.5\linewidth]{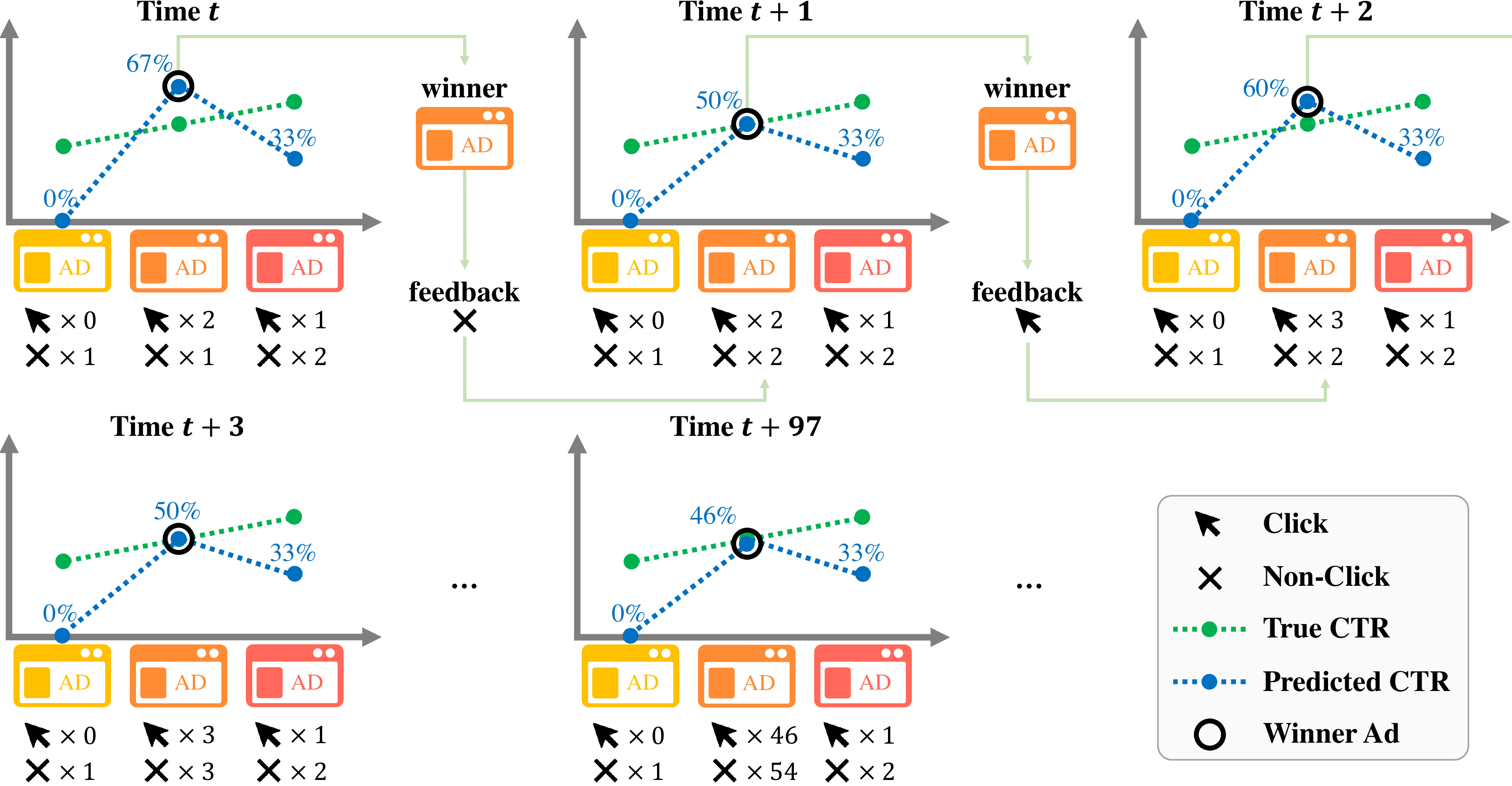}
  \vspace{-.5cm}
  \caption{
    An illustration of the issue of \textit{lack of exploration}. Assume three Ads are bid equally. The training data (i.e., feedback) are listed in each time step. The blue and green dotted lines show the predicted and true CTRs, respectively. Without exploration, the system could consistently display the middle Ad due to its highest CTR prediction. The underestimation of the other two Ads has little chance to be corrected, thus leading to suboptimality. Best viewed in color.
  }
  \label{fig:example}
  \Description{}
  \vspace{-.5cm}
\end{figure}

However, we point out an inherent inconsistency that the distribution of training samples for the CTR model are in fact largely influenced by the ranking of predicted eCPMs, which are in turn interfered by the CTR model itself.
This contradicts the assumption behind supervised learning that the data are i.i.d. from a fixed data-generating distribution $p_{_{\mathcal{X}\times\mathcal{Y}}}$ and can further lead to lack of \emph{exploration}:
only those Ads that the models believe to
have high eCPMs
are prone to be displayed and then collected into future data.
This is likely to result in inaccurate predictions for those Ads that have rarely or never been displayed, which further violates the assumption of perfect eCPM estimations behind the Ad-ranking strategy.
Moreover, since the CTR models are usually regularly retrained to incorporate the latest data, it potentially forms a pernicious \emph{feedback loop}~\cite{chaney2018algorithmic} (see \figref{loop} and a concrete example in \figref{example}), which can gradually lead to consistent inaccuracy in predictive performance, suboptimal auction mechanisms,
and even unintended social and economic consequences~\cite{chander2016racist} such as price discrimination~\cite{myerson1981optimal}.
Such self-interference of CTR models motivate us to move towards a contextual bandit formulation of advertising.





\vspace{-.2cm}
\section{A Contextual Bandit Formulation}
\label{sec:cb}

We now present a joint formulation of the CTR prediction and the Ad-ranking strategy from a contextual bandit~\cite{li2010contextual} perspective, which explicitly takes the sequential nature into consideration
to overcome the issue of lack of exploration discussed in \secref{back:drawback}.




Let $\mathcal{A}$ be the set of available Ads.
Formally, we consider advertising as a decision problem that takes a \emph{user feature} and an \emph{Env feature} $(x_{_U},x_{_E})\in\mathcal{X}_{_U}\times\mathcal{X}_{_E}$ as input and outputs an ordering of $\mathcal{A}$, which describes the Ad-ranking process before an impression.
The inputs $(x_{_U},x_{_E})$ here are assumed to be drawn i.i.d. from a fixed distribution $p_{_{\mathcal{X}_{_U}\times\mathcal{X}_{_E}}}$, which can be viewed as the statistical description of online traffic.
We define scoring functions $s:\mathcal{A}\times\mathcal{X}_{_U}\times\mathcal{X}_{_E}\rightarrow\mathbb{R}$ that score each Ad $\mathfrak{a}\in\mathcal{A}$ given $(x_{_U},x_{_E})$.
Clearly, any scoring function defines an ordering of $\mathcal{A}$, and thus corresponds to an Ad-ranking strategy.
For example, the strategy that ranks Ads by predicted eCPMs can be represented by the following scoring function:
\begin{equation}\label{eqn:ecpm-greedy}
  s_{\textrm{eCPM}}(\mathfrak{a};x_{_U},x_{_E}) \defeq \sigma(\hat{\slf}(x_{_U},x_{_A}(\mathfrak{a}),x_{_E})) \cdot \mathrm{Bid}(\mathfrak{a}),
\end{equation}
where $x_{_A}(\mathfrak{a})$ and $\mathrm{Bid}(\mathfrak{a})$ are the \emph{Ad feature} and the \emph{bid price} of the Ad $\mathfrak{a}$, respectively.
In the rest of this section, we assume only one Ad is displayed each time, for simplicity. In this case, the winner Ad $\mathfrak{a}$ is the top ranking one, i.e., $\mathfrak{a}=\argmax_{\mathfrak{a}'\in\mathcal{A}}s(\mathfrak{a}';x_{_U},x_{_E})$.

We further define stochastic Ad-ranking strategies, which hold a distribution $p(s)$ over scoring functions and output orderings of $\mathcal{A}$ with randomly sampled scoring functions from $p(s)$.
Note that the deterministic Ad-ranking strategies such as \eqn{ecpm-greedy} also fall into this formulation, where all the probability mass of $p(s)$ is assigned to one single scoring function. We therefore consider stochastic Ad-ranking strategies in the following, without loss of generality.
With a slight abuse of notation, we also denote a stochastic strategy by $s$ and the winner Ad by $\mathfrak{a}\sim s(\mathcal{A}|x_{_U},x_{_E})$, where $s(\mathcal{A}|x_{_U},x_{_E})$ is the distribution of winner Ads induced by $s$.

In advertising systems, we can only observe a stochastic feedback $y$ for the displayed Ad $\mathfrak{a}\sim s(\mathcal{A}|x_{_U},x_{_E})$.
Specifically, positive feedback $y=1$ and negative feedback $y=0$ are assumed to be drawn from an unknown conditional distribution $p_{_{\mathcal{Y}|\mathcal{X}}}$, i.e., $y\sim p_{_{\mathcal{Y}|\mathcal{X}}}(y|x_{_U},x_{_A}(\mathfrak{a}),x_{_E})$. 
Since the feedback $y$ only takes values of $1$ and $0$ in our formulation, the CTR is just the expectation $\E_{p_{_{\mathcal{Y}|\mathcal{X}}}(y|x_{_U},x_{_A}(\mathfrak{a}),x_{_E})}[y]$ which we denote by $\eyx{}$ for simplicity.

We then define the \emph{expected utility} of a stochastic strategy $s$ as:
\begin{equation}\label{eqn:utility}
  R(s) = \E_{(x_{_U},x_{_E})\sim p_{_{\mathcal{X}_{_U}\times\mathcal{X}_{_E}}}}\E_{\mathfrak{a}\sim s(\mathcal{A}|x_{_U},x_{_E})}\left(\eyx{}\cdot\mathrm{Bid}(\mathfrak{a})\right),
\end{equation}
which can be understood as the ``social welfare per impression'' provided that all advertisers bid truthfully.
The problem of advertising is then transformed to maximizing the expected utility $R(s)$ w.r.t. the stochastic strategy $s$.
A clear bound on $R(s)$ is as follows:
\begin{align}\label{eqn:ineq}
  R(s) & \leq \E_{(x_{_U},x_{_E})\sim p_{_{\mathcal{X}_{_U}\times\mathcal{X}_{_E}}}}\max_{\mathfrak{a}\in\mathcal{A}}\left(\eyx{}\cdot\mathrm{Bid}(\mathfrak{a})\right)\defeq R(s^*),
\end{align}
where $s^*$ is the optimal Ad-ranking strategy:
\begin{equation}\label{eqn:optmstr}
s^*(\mathfrak{a};x_{_U},x_{_E}) \defeq \eyx{}\cdot\mathrm{Bid}(\mathfrak{a}).
\end{equation}
The interpretation of \eqnt{ineq}{optmstr} is very intuitive:
The utility (social welfare) is maximized when the strategy is able to identify and display the Ads with highest true eCPM given the user and environment, which exactly implements the GSP auctions.

However, since the true CTRs/eCPMs are unknown, optimizing the strategy $s$ is essentially equivalent to maximizing the black-box eCPM function $\E[y|(x_{_U},x_{_A}(\mathfrak{a}),x_{_E})]\cdot\mathrm{Bid}(\mathfrak{a})$ w.r.t. $\mathfrak{a}\in\mathcal{A}$, which can be very difficult.
To make matters worse, it is even unreliable to get an approximate solution based on historical data,
since the optimal Ads may have never been displayed.

Fortunately, we can make trails (i.e., to display the Ads and get user feedback) to learn more information about the true CTRs and improve the strategy.
Specifically, given the \emph{user feature} and \emph{Env feature} $(x_{_U}^{(t)},x_{_E}^{(t)})$ at time $t$, the system is able to determine and display the winner Ad $\mathfrak{a}^{(t)}\sim s^{(t)}(\mathcal{A}|x_{_U}^{(t)},x_{_E}^{(t)})$ according to the strategy $s^{(t)}$, then receives feedback $y^{(t)}\sim p_{_{\mathcal{Y}|\mathcal{X}}}(y|x_{_U}^{(t)},x_{_A}(\mathfrak{a}^{(t)}),x_{_E}^{(t)})$,
and finally updates the strategy to $s^{(t+1)}$ based on the observed data $\data_{t}\defeq\{(x^{(i)},y^{(i)})\}_{i=1}^t$.
After $T$ rounds, the total utility is $\sum_{t=1}^T(y^{(t)}\cdot\mathrm{Bid}(\mathfrak{a}^{(t)}))$,
which is the objective we aim to maximize through the sequential decision-making process.

Intuitively, the strategy should favor the Ads with high eCPM predictions as well as the Ads that reveal more information about the CTR function.
This is known as the \emph{exploration-exploitation} trade-off and has been well discussed in previous studies of \textit{contextual bandit} problems~\cite{li2010contextual,chapelle2011empirical,mcinerney2018explore}.
However, in the advertising/recommendation community, 
most existing methods impose strong assumptions on the unknown CTR functions, limiting their flexibility.
Noticing that the key is to establish the confidence bounds or uncertainty estimations for the unknown functions, we propose a novel method for learning deep CTR models that enables uncertainty estimations, and show how it can be combined with bandit algorithms to obtain new practical Ad-ranking strategies in \secref{method}.


\vspace{-.2cm}
\section{Deep Uncertainty-Aware Learning}
\label{sec:method}

In this section, we present a novel \emph{Deep Uncertainty-Aware Learning} (DUAL) method for CTR prediction which enables uncertainty estimations and propose \emph{DUAL-based Ad-ranking strategies} for the contextual bandit problems of advertising described in \secref{cb}.

\vspace{-.2cm}
\subsection{DUAL for CTR Prediction}
\label{sec:method:svgp}

Training CTR prediction models is essentially learning the unknown CTR function $\eyx{}$ w.r.t. the feature vectors $x\in\mathcal{X}$ from the observed data $\data_t$.
Here we equivalently focus on the logit function, defined as $f(x)\defeq\sigma^{-1}(\eyx{})$, which has a range of $\mathbb{R}$ and is thus convenient for analysis.
Most existing deep learning-based CTR models approximate $f$ with a parameterized neural network $\hat{h}$ (as described in \secref{back:ctr}), which gives point predictions of CTR with no measure of uncertainty.
Instead, under a Bayesian perspective of function learning, we impose a prior distribution $p(f)$ to the logit function $f$ and (approximately) infer the posterior distribution $p(f|\data_t)$ over functions as the learned model, which gives distributional CTR predictions and thus can provide uncertainty estimations along with the predictions.

The prior distribution expresses our prior belief about the logit function $f$.
A common choice of prior distributions over functions is the Gaussian process (GP)~\cite{rasmussen2003gaussian}, which is favored due to its high flexibility for function learning.
Below, we describe how we obtain CTR predictions and uncertainties with GPs.

\noindent
\textbf{Gaussian Processes for CTR Prediction.}
We assume a GP prior distribution over $f$, i.e., $f\sim\mathcal{GP}(m(x), k(x,x'))$, where $m(x)=\E[f(x)]$ is the mean function and $k(x,x')=\mathrm{Cov}[f(x), f(x')]$ is the covariance kernel function.
For a thorough introduction of GPs, we refer to \citet{rasmussen2003gaussian}.

We aim to estimate the logit $f^*=f(x^*)$ as well as its uncertainty at any feature point $x^*\in\mathcal{X}$, given the observed data samples $\data_{t}=\{(x^{(i)},y^{(i)})\}_{i=1}^t$ at time $t$.
Let $\mathbf{X}_{t}=[x^{(1)},\cdots,x^{(t)}]^\top\in\mathbb{R}^{t\times d}$ be the matrix of the observed feature vectors, $\mathbf{f}_{t}=f(\mathbf{X}_{t})\in\mathbb{R}^t$ be the vector of the corresponding function values, and $\mathbf{y}_{t}=[y^{(1)},\cdots,y^{(t)}]^\top\in\{0,1\}^t$ be the vector of user feedback.
Each feedback $y$ is viewed as a noisy measurement of the logit $f(x)$, generated from a Bernoulli likelihood model $p(y|f(x))=\mathrm{Ber}(y;\sigma(f(x)))$.
Then the joint likelihood of $\mathbf{y}_{t}$, $\mathbf{f}_{t}$ and the logit value $f^*$ is:
\begin{equation}
  \!\!\!p(\mathbf{y}_{t},\mathbf{f}_{t},f^*)=p(\mathbf{y}_{t}|\mathbf{f}_{t})p(\mathbf{f}_{t},f^*)=p(\mathbf{f}_{t},f^*)\prod{}_{i=1}^t p(y^{(i)}|f^{(i)}),
\end{equation}
where $p(\mathbf{f}_{t},f^*)$ is a multivariate Gaussian distribution defined by the GP prior:
\vspace*{-.1cm}
\begin{equation}
  p(\mathbf{f}_{t},f^*)=\N\left(
    \begin{array}{c|}
    \mathbf{f}_{t} \\
    f^*
    \end{array}
    \begin{bmatrix}
      m(\mathbf{X}_{t}) \\
      m(x^*)
    \end{bmatrix},
    \begin{bmatrix}
        k(\mathbf{X}_{t},\mathbf{X}_{t}) & k(\mathbf{X}_{t},x^*)\\
        k(x^*,\mathbf{X}_{t}) & k(x^*,x^*)\\
    \end{bmatrix}\right),
\end{equation}
where $k(\mathbf{X}_{t},\mathbf{X}_{t})$ is a $t\times t$ matrix with the element $[k(\mathbf{X}_{t},\mathbf{X}_{t})]_{ij}$ being $k(x^{(i)},x^{(j)})$ and
$k(\mathbf{X}_{t},x^*)$ is defined similarly.
Then the posterior distribution of $f^*$ given $\data_{t}$ can be evaluated as:
\begin{equation}\label{eqn:exact-inf}
  \!\!p(f^*|\mathbf{y}_{t})\!=\!\frac{\int p(\mathbf{y}_{t},\mathbf{f}_{t},f^*) \mathrm{d}\mathbf{f}_{t}}{p(\mathbf{y}_{t})}\!=\!\frac{1}{p(\mathbf{y}_{t})}\int p(\mathbf{y}_{t}|\mathbf{f}_{t})p(\mathbf{f}_{t},f^*) \mathrm{d}\mathbf{f}_{t},
\end{equation}
which is the distributional prediction we desire.
However, the posterior is generally intractable since the Bernoulli likelihood is not conjugate to the GP prior.
Moreover, since the feature vectors are usually extremely sparse and high-dimensional in advertising systems, it is also challenging to design a proper kernel function that exploits the structural information of the feature space $\mathcal{X}$.
Inspired by~\citet{wilson2016deep,wilson2016stochastic},
we propose to address these challenges by 1) incorporating the deep structures into base kernel functions and 2) adopting sparse variational approximate inference for GPs~\cite{titsias2009variational}, so that we can benefit from both the expressive power of deep models and the ability of GPs to estimate uncertainties.


\noindent
\textbf{Deep Kernels.}
Existing deep CTR models tackle the sparsity and high-dimensionality with embeddings and learn high-level abstraction of input vectors with neural networks~\cite{zhou2018deep,guo2017deepfm}.
These models can be summarized as first extracting a compact feature representation $\phi(x;\theta)\in\mathbb{R}^{d'}$ from the input vector $x\in\mathcal{X}\subset\mathbb{R}^{d}$ with a nonlinear mapping $\phi(\cdot;\theta)$ parameterized by $\theta$,
and then applying a simple logistic regression to the compact feature $\phi(x;\theta)$.
We propose to build up deep kernels by combining base kernels with deep nonlinear mappings.
Specifically, given a deep nonlinear mapping $\phi(\cdot;\theta)$ and a base kernel $k_0(x,x')$,
such as the RBF kernel $k_0(x,x')=a^2\cdot\exp(-\frac{\|x-x'\|^2}{2l^2})$, the composed kernel is defined as:
\begin{equation}
  k(x,x')=k_0(\phi(x;\theta), \phi(x';\theta)).
\end{equation}
It is clear that this combination is compatible with any structures of deep CTR models and any valid base kernel functions.


\noindent
\textbf{Sparse Variational Approximation of GP Inference.}
Inference from \eqn{exact-inf} is generally very hard due to the non-conjugacy and the scalability limitation in GPs.
Sparse variational GP (SVGP) methods~\cite{hensman2015scalable,titsias2009variational,hensman2013gaussian}
have shown promise in addressing these difficulties.
The core idea of SVGP methods is to approximate the posterior with a bunch of learned inducing points that best summarize the observed data.
We propose to adopt SVGP methods for efficient and scalable approximate inference of the CTR as well as its uncertainty.

Specifically, let $\mathbf{Z}=[z_1,\cdots,z_M]^\top\in\mathbb{R}^{M\times d}$ be a set of $M$ parameterized vectors in $\mathcal{X}$, called the inducing points, and let $\mathbf{u}=f(\mathbf{Z})$. Then the joint distribution of $\mathbf{u}$, $\mathbf{f}_t$, $f^*$ and $\mathbf{y}_{t}$ given by the GP is:
\begin{equation}\label{eqn:augjoint}
  p(\mathbf{y}_{t},\mathbf{f}_{t},f^*,\mathbf{u})=p(\mathbf{f}_{t},f^*|\mathbf{u})p(\mathbf{u})\prod{}_{i=1}^t p(y^{(i)}|f^{(i)}),
\end{equation}
where both $p(\mathbf{u})=\N(m(\mathbf{Z}), k(\mathbf{Z},\mathbf{Z}))$ and $p(\mathbf{f}_{t},f^*|\mathbf{u})$ are Gaussian according to the property of GPs.
Then the posterior inference $p(f^*|\mathbf{y}_{t})$ can be done in the augmented model \eqn{augjoint} and
transformed into a variational inference problem,
which approximates the posterior $p(\mathbf{f}_{t},f^*,\mathbf{u}|\mathbf{y}_{t})$ with a variational distribution $q(\mathbf{f}_{t},f^*,\mathbf{u})$ by minimizing $\KL[q(\mathbf{f}_{t},f^*,\mathbf{u})\|p(\mathbf{f}_{t},f^*,\mathbf{u}|\mathbf{y}_{t})]$.

A common choice of the variational distribution is $q(\mathbf{f}_{t},f^*,\mathbf{u})\defeq q(\mathbf{u})p(\mathbf{f}_{t},f^*|\mathbf{u})$, where $q(\mathbf{u})=\N(\mathbf{v},\mathbf{S})$ is assumed Gaussian.
The mean vector $\mathbf{v}$ and the covariance matrix $\mathbf{S}$ of $q(\mathbf{u})$ and the inducing points $\mathbf{Z}$ are all variational parameters to learn.
In this case, the posterior distribution of $f^*$, i.e. $p(f^*|\mathbf{y}_{t})$, can be approximated as:
\begin{equation}
\begin{aligned}\label{eqn:posterior}
  p(f^*&|\mathbf{y}_{t})=\int p(\mathbf{f}_{t},f^*,\mathbf{u}|\mathbf{y}_{t}) \mathrm{d} \mathbf{f}_{t} \mathrm{d} \mathbf{u}
  \approx \int q(\mathbf{f}_{t},f^*,\mathbf{u}) \mathrm{d} \mathbf{f}_{t} \mathrm{d} \mathbf{u}\\
  & =\int p(f^*|\mathbf{u})q(\mathbf{u}) \mathrm{d} \mathbf{u} \defeq q(f^*;\mathbf{v},\mathbf{S}) = \N(f^*|\mu^*,\Sigma^*),
\end{aligned}
\end{equation}
in which $\mathbf{f}_{t}$ and $\mathbf{u}$ are analytically integrated out and we reach a Gaussian marginal variational posterior $q(f^*;\mathbf{v},\mathbf{S})=\N(f^*|\mu^*,\Sigma^*)$.
It can be shown~\cite{salimbeni2017doubly} that the mean and variance are:
\begin{equation}\label{eqn:inference}
  \begin{aligned}
  &\mu^*=m(x^*)+k(x^*,\mathbf{Z})\mathbf{K}_{\mathbf{uu}}^{-1}(\mathbf{v}-m(\mathbf{Z}))\\
  \Sigma^*={}&k(x^*,x^*)- k(x^*,\mathbf{Z})\mathbf{K}_{\mathbf{uu}}^{-1}\left(\mathbf{K}_{\mathbf{uu}}-\mathbf{S}\right)\mathbf{K}_{\mathbf{uu}}^{-1}k(x^*,\mathbf{Z})^\top,
  \end{aligned}
\end{equation}
where $k(x^*,\mathbf{Z})$ is an $M$-dimensional vector with $[k(x^*,\mathbf{Z})]_m=k(x^*,z_m)$ and we denote $\mathbf{K}_{\mathbf{uu}}=k(\mathbf{Z},\mathbf{Z})$ for simplicity.

Finally, the entire model is learned by minimizing the KL divergence $\KL[q(\mathbf{f}_{t},f^*,\mathbf{u})\|p(\mathbf{f}_{t},f^*,\mathbf{u}|\mathbf{y}_{t})]$ w.r.t. both the variational parameters $\{\mathbf{Z},\mathbf{v},\mathbf{S}\}$ and the deep kernel parameters $\theta$,
which is equivalent to maximize the evidence lower bound (ELBO):
\begin{equation}\label{eqn:elbo}
  \!\!\!\!\!\mathcal{L}(\data_t)\!=\!
  \sum{}_{i=1}^t\E_{q(f^{(i)};\mathbf{v},\mathbf{S})}[\log p(y^{(i)}|f^{(i)})] - \KL[q(\mathbf{u})\|p(\mathbf{u})].
\end{equation}
The summation form over data in \eqn{elbo} allows us to perform mini-batch training on large datasets. We name the proposed method for CTR prediction as \emph{Deep Uncertainty-Aware Learning} (DUAL).

\vspace{-.2cm}
\subsection{DUAL-Based Ad-Ranking Strategies}
\label{sec:method:adrank}
We present two Ad-ranking strategies based on DUAL, in which we combine the uncertainties provided by DUAL with two popular bandit algorithms: Upper Confidence Bounds (UCB)~\cite{auer2002using} and Thompson Sampling (TS)~\cite{chapelle2011empirical}, which are widely studied for decades.

\noindent
\textbf{DUAL-UCB.}
DUAL-UCB scores the Ads with the upper confidence bounds on the eCPMs:
\begin{equation}\label{eqn:dualucb}
  s_{\mathrm{UCB}}(\mathfrak{a};x_{_U},x_{_E}) \defeq \sigma\left(\mu(\mathfrak{a}) + \kappa\cdot \sqrt{\Sigma(\mathfrak{a})}\right) \cdot \mathrm{Bid}(\mathfrak{a}),
\end{equation}
where $\mu(\mathfrak{a})$ and $\Sigma(\mathfrak{a})$ denotes the mean and variance of the posterior logit value $f(x^*)$ at $x^*=[x_{_U}^\top,x_{_A}(\mathfrak{a})^\top,x_{_E}^\top]^\top$, and $\kappa$ is a hyperparameter balancing the exploration-exploitation trade-off.

\noindent
\textbf{DUAL-TS.} DUAL-TS is a stochastic strategy that scores Ads following the posterior eCPM distribution:\\[-.3cm]
\begin{equation}\label{eqn:dualts}
  s_{\mathrm{TS}}(\mathfrak{a};x_{_U},x_{_E}) \defeq \sigma(\hat{f}) \cdot \mathrm{Bid}(\mathfrak{a}),\quad \hat{f}\sim\N(\mu(\mathfrak{a}),\Sigma(\mathfrak{a})),
\end{equation}
  where $\hat{f}$ is randomly sampled from the posterior in \eqn{posterior}.

Intuitively, both DUAL-UCB and DUAL-TS can balance exploration against exploitation, since they both encourage the system to explore new Ads or Ads with few data samples, which are expected to have high posterior uncertainties in their eCPM estimations.

Meanwhile, we also describe a greedy strategy:

\noindent
\textbf{DUAL-Greedy.} DUAL-Greedy scores the Ads with predicted eCPMs while ignoring the uncertainty:
  \begin{equation}\label{eqn:dualgd}
    s_{\mathrm{Greedy}}(\mathfrak{a};x_{_U},x_{_E}) \defeq \sigma\left(\mu(\mathfrak{a}) \right) \cdot \mathrm{Bid}(\mathfrak{a}).
  \end{equation}
Note that it is greedy w.r.t. the maximum a posteriori (MAP) estimates $\mu(\mathfrak{a})$, in which the prior still plays an important role (See \eqn{inference}). If we choose a GP prior with a constant mean function $m(x)=C$ and set a higher constant $C$, the predictions to new Ads or Ads with few samples will also be higher.
Under such circumstances, DUAL-Greedy will be inclined to display these Ads even if it ignores uncertainty.
This suggests that with a properly specified prior, DUAL-Greedy also has the potential to conduct exploration.

It is also worth noting that the common paradigm of advertising, i.e., training a deep CTR model in the purely supervised learning manner (as described in \secref{back:ctr}) and ranking Ads by predicted eCPMs (as in \eqn{ecpm-greedy}), can be viewed as a special case of the greedy strategy, to which we refer as the \textbf{DNN-Greedy} strategy.
This can be argued by comparing \eqn{ecpm-greedy} with \eqn{dualgd} and the fact that $\hat{h}$ corresponds to a MAP estimate of the logit function $f$ in the nonparametric sense if we choose an uninformative prior~\cite{murphy2012machine} over $f$.
Therefore, we conjecture that DUAL-Greedy can also improve over DNN-Greedy as we can choose a more appropriate and informative prior, which is empirically verified in \secref{exp:explore}. 

\vspace{-.2cm}
\subsection{Practical Techniques for DUAL}
\label{sec:method:prac}

We describe several practical techniques for applying DUAL in real-world CTR prediction tasks.

\noindent
\textbf{Inducing Points \& Regularization in Hidden Space.}
Noticing that the inducing points $\mathbf{Z}$ always appear in the deep kernel, i.e., $k(\cdot,\mathbf{Z})=k_0(\phi(\cdot;\theta),\phi(\mathbf{Z};\theta))$, we directly parameterize $\mathbf{Z}'\defeq\phi(\mathbf{Z};\theta)$ $\in\mathbb{R}^{M\times d'}$ in the hidden space instead of $\mathbf{Z}\in\mathbb{R}^{M\times d}$ in the input space, which extremely reduces the number of parameters to learn.

However, this raises a new issue of training when using kernels like RBF as the base kernel.
Specifically, with a randomly initialized $\theta$ and $\mathbf{Z}'$, it is very likely to happen that $\phi(\cdot;\theta)$ maps input data $\mathbf{X}_t$ to somewhere far from $\mathbf{Z}'$ in the hidden space, which can cause gradient saturation and failure in learning.
Inspired by prior work~\cite{hensman2015scalable} that initializes the inducing points using $K$-means clustering, we propose to regularize the inducing points $\mathbf{Z}'$ and the mapping parameters $\theta$ with a clustering loss $\mathcal{R}(\data_t)$:\\[-.2cm]
\begin{equation}
  \mathcal{R}(\data_t) = \frac{1}{t}\sum{}_{i=1}^t \left\|\phi(x^{(i)};\theta)-z_{c_i}'\right\|^2,
\end{equation}
where $c_i=\argmin_{m\in[M]}\|z_m'-\phi(x^{(i)};\theta)\|^2$ is the index of nearest inducing point of $\phi(x^{(i)};\theta)$. This regularization encourages the inducing points to track the data in the hidden space while the mapping $\theta$ is deforming during training.

\noindent
\textbf{Temperature Control.} We incorporate a \emph{temperature} $\tau$ to control the strength of the prior, which is a common practice in Bayesian inference~\cite{wenzel2020good}. Specifically, we modified the ELBO in \eqn{elbo} to:\\[-.3cm]
\begin{equation}\label{eqn:fff}
  \!\!\!\!\!\mathcal{L}_\tau(\data_t)=\frac{1}{t}\!\sum_{i=1}^t\!\Em_{q(f^{(i)};\mathbf{v},\mathbf{S})}[\log p(y^{(i)}|f^{(i)})] - \tau\cdot\KL[q(\mathbf{u})\|p(\mathbf{u})].
\end{equation}
Then, the final loss function for practical training of DUAL is:\\[-.2cm]
\begin{equation}
  \mathcal{O}(\theta,\mathbf{Z}',\mathbf{v},\mathbf{S};\tau,\lambda) = - \mathcal{L}_\tau(\data_t) + \lambda \cdot \mathcal{R}(\data_t),
\end{equation}
where both $\tau$ and $\lambda$ are tunable hyperparameters.

\vspace{-.2cm}
\subsection{Computational and Memory Complexity}
\label{sec:method:complexity}
\noindent
In addition to the deep mapping parameters $\theta$, DUAL introduces variational parameters $\mathbf{Z}'\in\mathbb{R}^{M\times d'}$, $\mathbf{v}\in\mathbb{R}^M$ and $\mathbf{S}\in\mathbb{R}^{M\times M}$. The extra memory complexity is therefore $O(Md'+M^2)$.
Note that one can use structured covariance approximations such as a diagonal covariance $\mathbf{S}=\mathrm{diag}(s_1,\cdots,s_M)$ to further reduce the parameters.

The terms of $\bm{\alpha}_1\defeq \mathbf{K}_{\mathbf{uu}}^{-1}(\mathbf{v}-m(\mathbf{Z}))$ and $\bm{\alpha}_2\defeq \mathbf{K}_{\mathbf{uu}}^{-1}\left(\mathbf{K}_{\mathbf{uu}}-\mathbf{S}\right)\mathbf{K}_{\mathbf{uu}}^{-1}$ in \eqn{inference} are independent with $x^*$ and thus can be precomputed before performing inference. 
During online inference, after computing the deep nonlinear mapping $\phi(x^*;\theta)$, we need to 1) calculate $\bm{\beta}\defeq k(x^*,\mathbf{Z})=k_0(\phi(x^*;\theta),\mathbf{Z}')$, which costs $O(Md')$ time; and 2) calculate $\bm{\beta}\bm{\alpha}_1$ and $\bm{\beta}\bm{\alpha}_2\bm{\beta}^\top$, which costs $O(M^2)$ time.
Therefore, the extra computational complexity of obtaining the predictive mean and variance is also $O(Md'+M^2)$, which is almost equivalent to the complexity of an additional $M$-dimensional layer in deep networks for a moderate $M$.
We also empirically observe that it cause minimal additional computational overhead in large deep CTR models.

\vspace{-.2cm}
\section{Related Work}\label{sec:relatedwork}

\noindent
\textbf{CTR Prediction.}
CTR prediction has proven to be an important personalization technique in modern recommendation/advertising systems.
Recently, deep learning approaches have been introduced in
CTR prediction
methods with great success~\cite{guo2017deepfm,he2017neural,cheng2016wide,zhou2018deep,zhou2019deep,feng2019deep}, due to their rich representation ability.
\citet{cheng2016wide} combine wide linear models and deep neural networks to enjoy both memorization and generalization abilities.
\citet{zhou2019deep} introduces recurrent units to model the evolving processes in user interests.
These works mainly focus on designing better model structures and learning more effective representations of user interests.
We instead propose a learning approach without specifying a concrete deep structure.
Our work can be viewed as a complementary approach to these existing deep CTR prediction methods,
in that it enables uncertainty estimation for \textit{arbitrary} deep CTR model structures.

\noindent
\textbf{Exploration-Exploitation Trade-off.}
The issue of exploration-exploitation trade-off in recommendation/advertising systems has been discussed for decades~\cite{balabanovic1998exploring,li2010contextual,vanchinathan2014explore,mcinerney2018explore}.
Early approaches~\cite{radlinski2008learning,sutton2018reinforcement} adopt context-free models, e.g., multi-armed bandit (MAB), in which no information is shared among users or items.
To address this challenge, recent research commonly considers the contextual bandit setting~\cite{li2010contextual,filippi2010parametric,krause2011contextual}, where users/items are related through features.
\citet{li2010contextual} propose LinUCB, which assumes a linear dependence between the feedback and the features.
Later, the dependence assumption is extended to logistic~\cite{chapelle2011empirical,dumitrascu2018pg} and generalized linear~\cite{li2017provably,filippi2010parametric}.
Although these methods greatly reduce the problem complexity, their assumptions are too strict and the flexibility is limited.
\citet{vanchinathan2014explore} propose to manage the exploration-exploitation trade-off in list recommendation problem based on Gaussian process bandits~\cite{krause2011contextual,srinivas2012information},
which is close to our idea.
However, our method is compatible with deep CTR models and thus can handle high-dimensional sparse data through embedding techniques,
which is not available in all above-mentioned methods.

\noindent
\textbf{Uncertainty Modeling in Deep Structures.}
Recently, there has been growing research interest in quantifying predictive uncertainty in deep neural networks~\cite{blundell2015weight,gal2016dropout,lakshminarayanan2017simple,guo2020deep}, which can be viewed as alternative ways to estimate uncertainties in deep CTR models.
Most of the existing work resorts to Bayesian neural networks (BNN),
where a prior distribution is imposed on the parameters of the neural network. The predictive uncertainty is then induced from the posterior distribution over parameters given the training data, which is obtained by variational inference~\cite{graves2011practical,blundell2015weight} or MCMC~\cite{gal2016dropout,welling2011bayesian} methods.
Another line of research focuses on ensemble learning~\cite{lakshminarayanan2017simple}, such as bagging~\cite{murphy2012machine}, where multiple networks are trained to produce uncertainty estimates.
However, most BNNs and ensemble methods cost significantly more inference time or storage, which is too expensive to practice in large-scale industrial applications.
On the contrary, our method only needs $O(Md'+M^2)$ additional computational and memory cost.
\citet{wilson2016deep,wilson2016stochastic} share a similar idea of combining GPs with deep kernels.
However, our work differs in two respects.
First, we parameterize the inducing points in the compact hidden space instead of the input space (which could be high-dimensional and sparse), thus greatly saving the number of parameters.
Second, their learning process includes two pre-training stages and a joint fine-tuning stage, which is difficult to handle in industrial practice.
We instead achieve end-to-end training with a clustering regularization.






\begin{figure*}[t]
  \vspace{-.1cm}
  \centering
  \begin{subfigure}[t]{0.19\textwidth}
    \includegraphics[width=\textwidth]{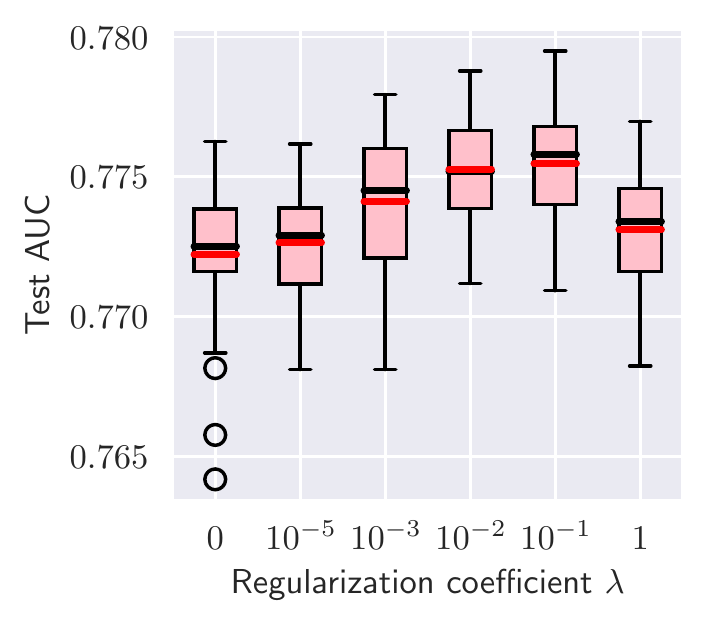} 
  \end{subfigure}
  \begin{subfigure}[t]{0.19\textwidth}
    \includegraphics[width=\textwidth]{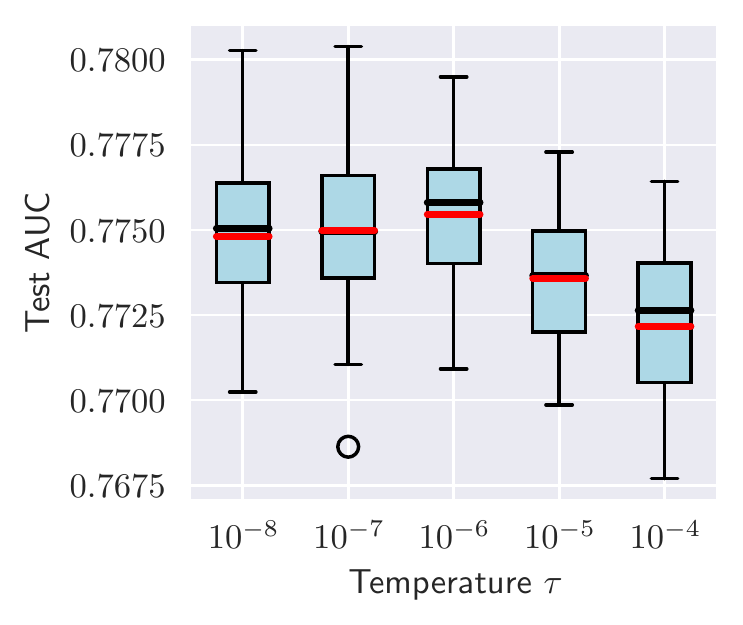} 
  \end{subfigure}
  \begin{subfigure}[t]{0.19\textwidth}
    \includegraphics[width=\textwidth]{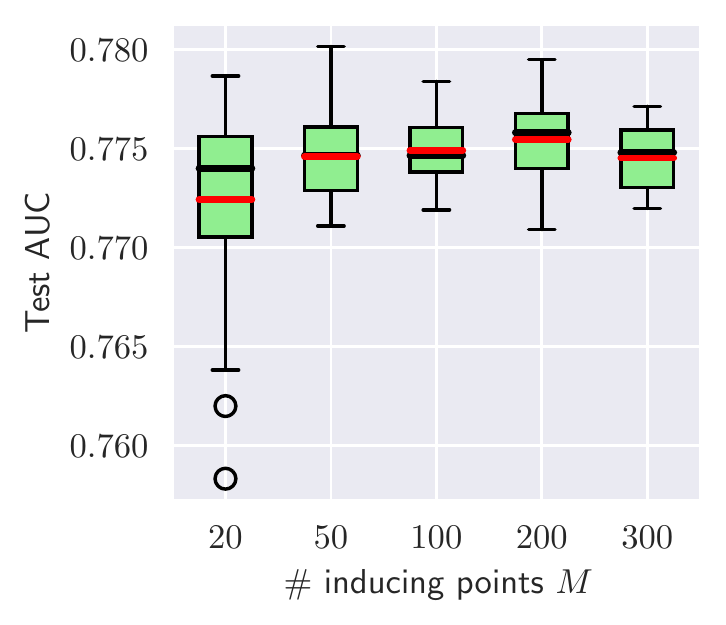} 
  \end{subfigure}
  \begin{subfigure}[t]{0.19\textwidth}
    \includegraphics[width=\textwidth]{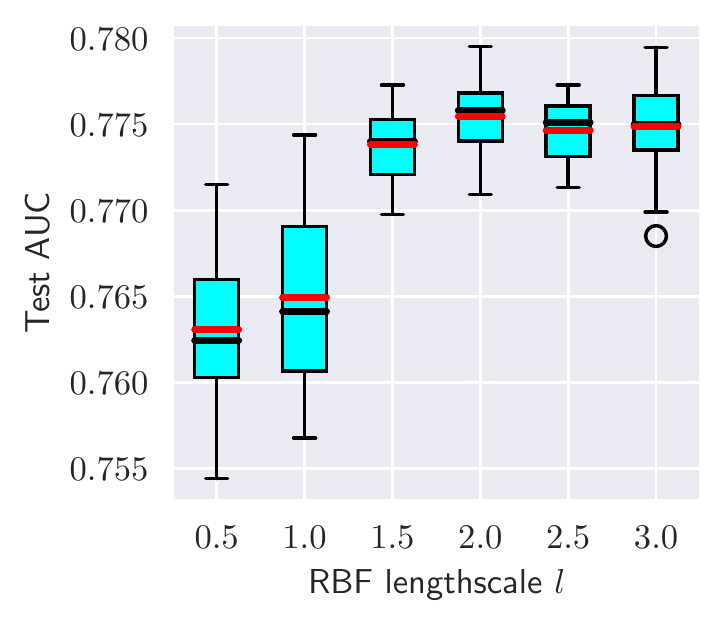} 
  \end{subfigure}
  \begin{subfigure}[t]{0.19\textwidth}
    \includegraphics[width=\textwidth]{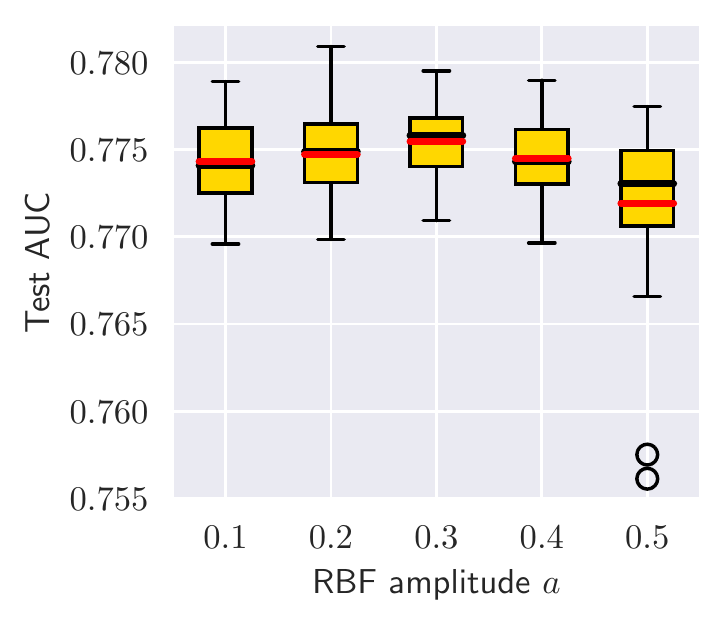} 
  \end{subfigure}
  \vspace{-.4cm}
  \caption{
    Sensitivity analysis of hyperparameters in DUAL.
    Each boxplot shows the results of 32 independent repeated runs on the Books dataset using the DNN architecture, where the red and black bars indicate the average and the median, respectively.
  }
  \label{fig:book-hyper}
  \Description{}
  \vspace{-.3cm}
\end{figure*}

\vspace{-.2cm}
\section{Experiments}\label{sec:exp}

We present experimental results of the proposed method on several public datasets to demonstrate its effectiveness.
We first examine the predictive performance of DUAL when it is combined with several state-of-the-art deep CTR models.
Then, we verify the efficacy of the proposed DUAL-based strategies in long-term social welfare maximization.
Finally, we share
the results and techniques of a two-week online A/B test in the Alibaba display advertising platform.

Throughout the experiments, we use RBF base kernels in DUAL, i.e., $k_0(x,x')=a^2\cdot\exp(-\frac{\|x-x'\|^2}{2l^2})$.
We adopt Adam~\cite{kingma2014adam} 
to optimize all model parameters.
Hyperparameters and detailed experimental settings will be specified later for each task.
Code is available at \blue{\url{https://github.com/duchao0726/DUAL}}.

\vspace{-.2cm}
\subsection{CTR Prediction}\label{sec:exp:ctr}

DUAL can be treated as a complementary component to deep CTR models which enables uncertainty estimation, as described in \secref{method:svgp}.
One common concern would be: does DUAL improve or hurt the predictive performance of the origin CTR models?
To answer this question, we combine several state-of-the-art deep CTR models with DUAL
and compare them with the original models.

We conduct the evaluation on two subsets of Amazon dataset~\cite{mcauley2015image}: Books and Electronics.
The data are preprocessed in the same way as in \citet{zhou2019deep}.
After preprocessing, Books consists of $1,086,120$ training samples and $121,216$ test samples.
Electronics consists of $346,110$ training samples and $38,696$ test samples.

The comparison is performed on five popular deep CTR models, including
\textbf{DNN}~\cite{zhou2018deep},
\textbf{Wide\&Deep}~\cite{cheng2016wide},
\textbf{PNN}~\cite{qu2016product},
\textbf{DIN}~\cite{zhou2018deep},
and \textbf{DIEN}~\cite{zhou2019deep}.
The network structures and the embedding layers of the five models are also set the same as in \citet{zhou2019deep}, 
which we refer to as the \emph{vanilla} models,
where all the models adopt a multilayer perceptron (MLP) with $200$-$80$-$2$ fully connected (FC) layers on top of their different compact representations of input features.
When learning with DUAL,
we treat these models as the deep nonlinear mappings, which map input fetures into a $2$-dimensional hidden space.
The inducing points in the hidden space are therefore of $2$ dimensions, i.e., $\mathbf{Z}'\in\mathbb{R}^{M\times 2}$.
We use diagonal covariances for the variational distribution, i.e., $\mathbf{S}=\mathrm{diag}(s_1,\cdots,s_M)$.
This results in an $O(M)$ ($4M$ in our case) variational parameters,
which is negligible compared to the number of weights in the deep neural networks and embedding layers.
The number of inducing points $M$ is set to $200$.
We choose $m(x)=0$ for the mean function of the GP prior.
The regularization coefficient is chosen from $\lambda=\{0.01,0.1,1.0\}$ and the temperature $\tau$ is set inversely proportional to the size of each dataset.
For the RBF base kernel, we choose the amplitude $a\in\{0.1, 0.3\}$ and the lengthscale $l\in[1.5, 3.0]$ manually.
We also provide a sensitivity analysis of hyperparameters.



\begin{figure}[tb]
  \centering
  \includegraphics[width=.95\linewidth]{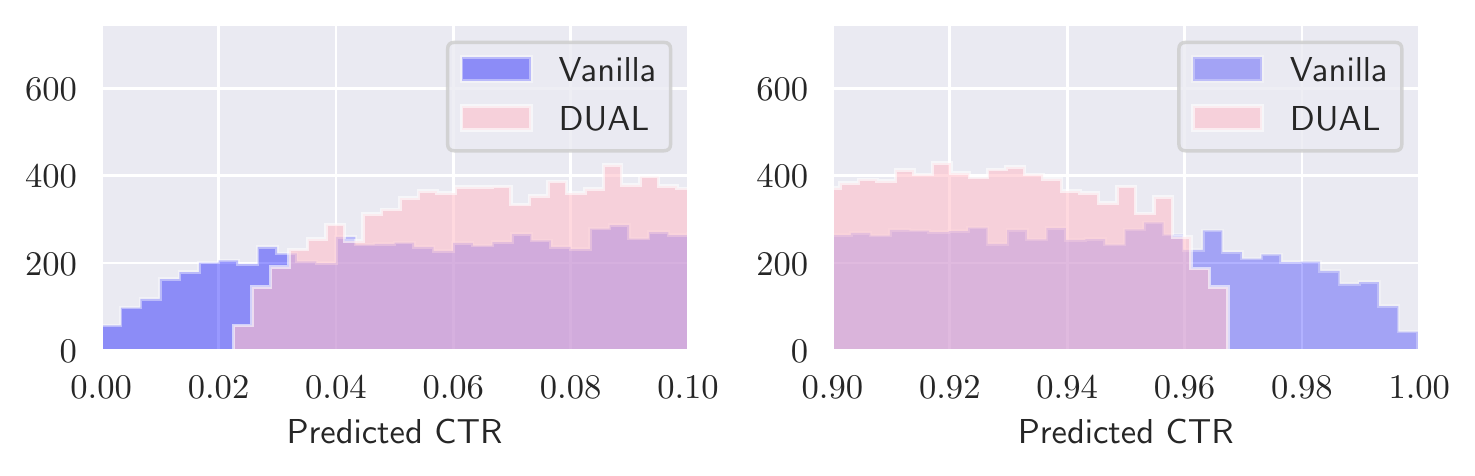} 
  \vspace{-.4cm}
  \caption{
    Histograms of the CTR predictions in the intervals of $[0,0.1]$ (Left) and $[0.9,1]$ (Right) on test data of Books from the vanilla DNN model and the DNN learned with DUAL.
  }
  \label{fig:book-dist}
  \Description{}
  \vspace{-.5cm}
\end{figure}

\tabref{ctr-book} and \tabref{ctr-electronic} show the comparison of the predictive results.
We report the average and standard deviation of test AUC estimates of $32$ repeated runs for each result.
For DUAL we use the predictive mean as the prediction.
We observe that learning these models with DUAL consistently produces better or comparable results.
We also test the vanilla models with larger MLPs ($200$-$90$-$2$ FC layers) which incorporate more parameters than DUAL and observe no improvement compared to the original ones.
This suggests that the improvements obtained from DUAL are not simply due to the slightly more parameters.
\figref{book-dist} shows the histograms of CTR predictions on test data from the vanilla DNN and the DNN learned with DUAL.
We observe that DUAL makes much less overconfident predictions,
which is as expected since the predictive mean of DUAL averages all possibilities. (Recall the example in \figref{example}. When only one negative sample is observed, the prediction of an ideal ERM is $0\%$, whereas DUAL will produce a positive predictive mean.)
It suggests that DUAL can be more robust to overfitting, benefiting from its Bayesian formalism and awareness of uncertainty.




\begin{table}[t]
  \caption{Predictive results (AUC) on Books}
  \label{table:ctr-book}
  \vspace{-.3cm}
  \begin{tabular}{lcc}
    \toprule
    Architecture & Vanilla & DUAL\\
    \midrule
    DNN~\cite{zhou2018deep} & $0.7682 \pm 0.0027$ & $\mathbf{0.7755 \pm 0.0020}$ \\
    Wide\&Deep~\cite{cheng2016wide} & $0.7729 \pm 0.0018$ & $\mathbf{0.7763 \pm 0.0019}$\\
    PNN~\cite{qu2016product} & $0.7801 \pm 0.0019$ & $\mathbf{0.7858 \pm 0.0021}$\\
    DIN~\cite{zhou2018deep} & $0.7885 \pm 0.0018$ & $\mathbf{0.7948 \pm 0.0019}$\\
    DIEN~\cite{zhou2019deep} & $0.8474 \pm 0.0033$ & $\mathbf{0.8492 \pm 0.0017}$\\
  \bottomrule
\end{tabular}
\vspace{-.6cm}
\end{table}

\noindent
\textbf{Sensitivity Analysis.}
We now investigate the influence of several hyperparameters on DUAL.
Here we consider the experiments of learning the DNN models with DUAL on the Book dataset.
\figref{book-hyper} shows the results produced with different hyperparameters.
First, we observe that the performance increases as the regularization coefficient $\lambda$ increases from $0$ to $10^{-1}$, which verifies the effectiveness of the proposed regularization technique for DUAL.
For the temperature $\tau$, we find that using the reciprocal of the number of data produces good results.
We observe that using more inducing points can improve the predictive performance.
However, we do not observe significant improvement with $M$ larger than $200$.
Finally, we find that the RBF base kernel has a considerable influence on the results. We thus manually choose the lengthscale $l$ and the amplitude $a$ of the RBF base kernel for each experiment.

\begin{table}[t]
  \caption{Predictive results (AUC) on Electronics}
  \label{table:ctr-electronic}
  \vspace{-.3cm}
  \begin{tabular}{lcc}
    \toprule
    Architecture & Vanilla & DUAL\\
    \midrule
    DNN~\cite{zhou2018deep} & $0.7413 \pm 0.0017$ & $\mathbf{0.7438 \pm 0.0017}$ \\
    Wide\&Deep~\cite{cheng2016wide} & $0.7441 \pm 0.0015$ & $\mathbf{0.7501 \pm 0.0013}$\\
    PNN~\cite{qu2016product} & $0.7541 \pm 0.0011$ & $\mathbf{0.7556 \pm 0.0008}$\\
    DIN~\cite{zhou2018deep} & $0.7593 \pm 0.0008$ & $\mathbf{0.7628 \pm 0.0008}$\\
    DIEN~\cite{zhou2019deep} & $0.7767 \pm 0.0023$ & $\mathbf{0.7824 \pm 0.0013}$\\
  \bottomrule
\end{tabular}
\vspace{-.2cm}
\end{table}

\begin{figure*}[t]
  \centering
  \begin{subfigure}[t]{0.33\textwidth}
    \includegraphics[width=\textwidth,height=.75\textwidth]{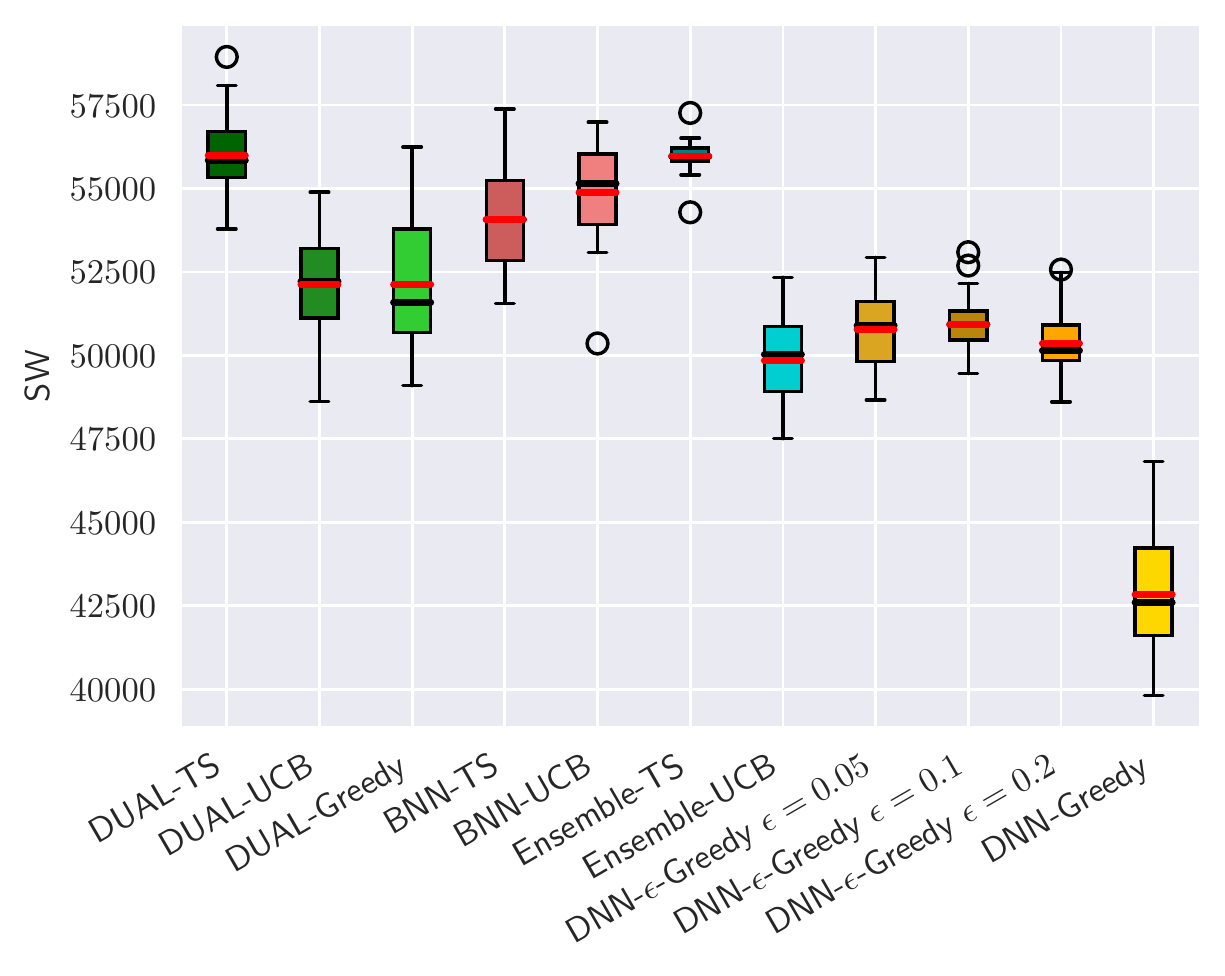} 
    \vspace{-.7cm}
    \caption{Final cumulative SW}\label{fig:r6b:swstd}
  \end{subfigure}
  \begin{subfigure}[t]{0.33\textwidth}
    \includegraphics[width=\textwidth,height=.75\textwidth]{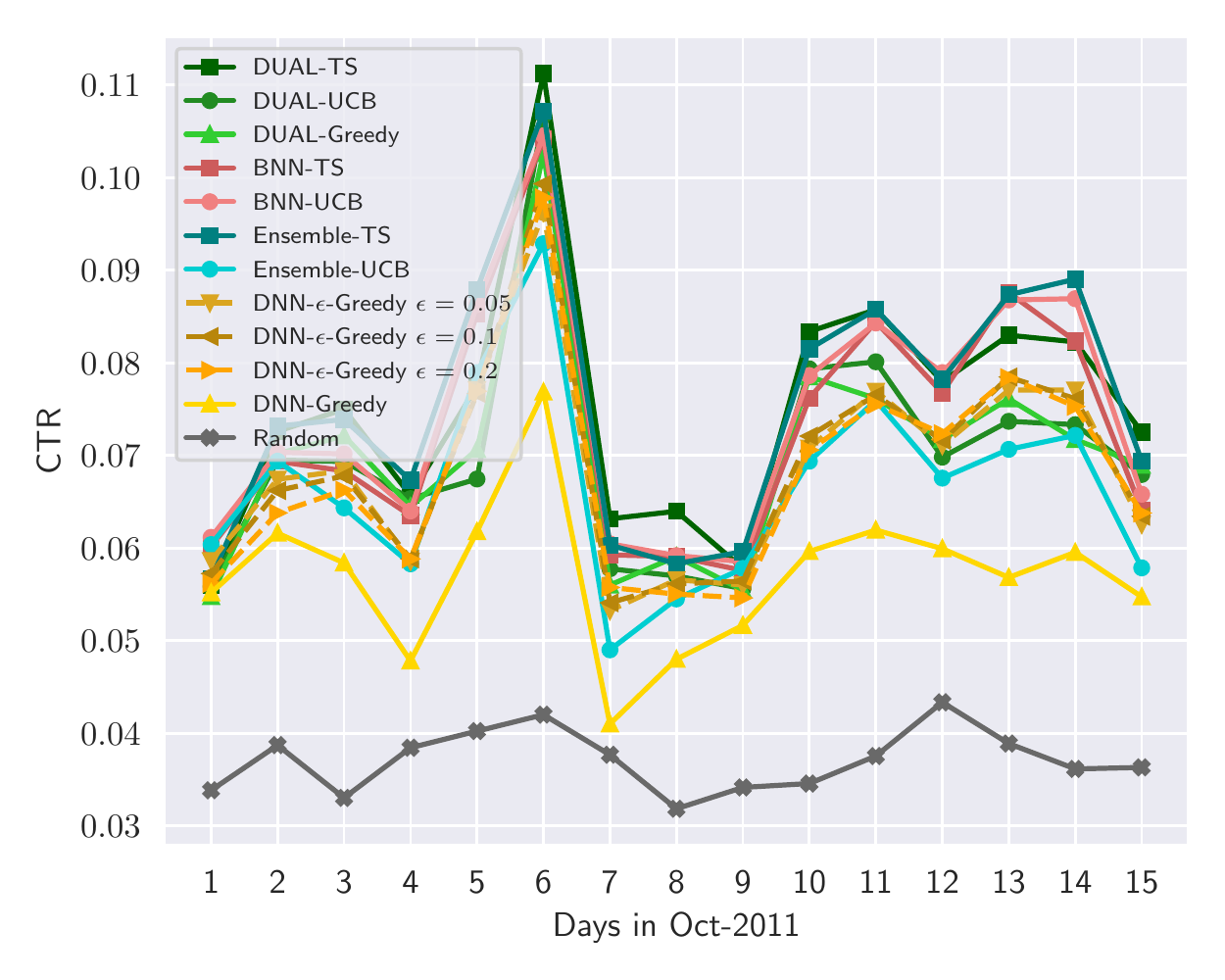} 
    \vspace{-.7cm}
    \caption{Daily CTRs}\label{fig:r6b:ctr}
  \end{subfigure}
  \begin{subfigure}[t]{0.33\textwidth}
    \includegraphics[width=\textwidth,height=.75\textwidth]{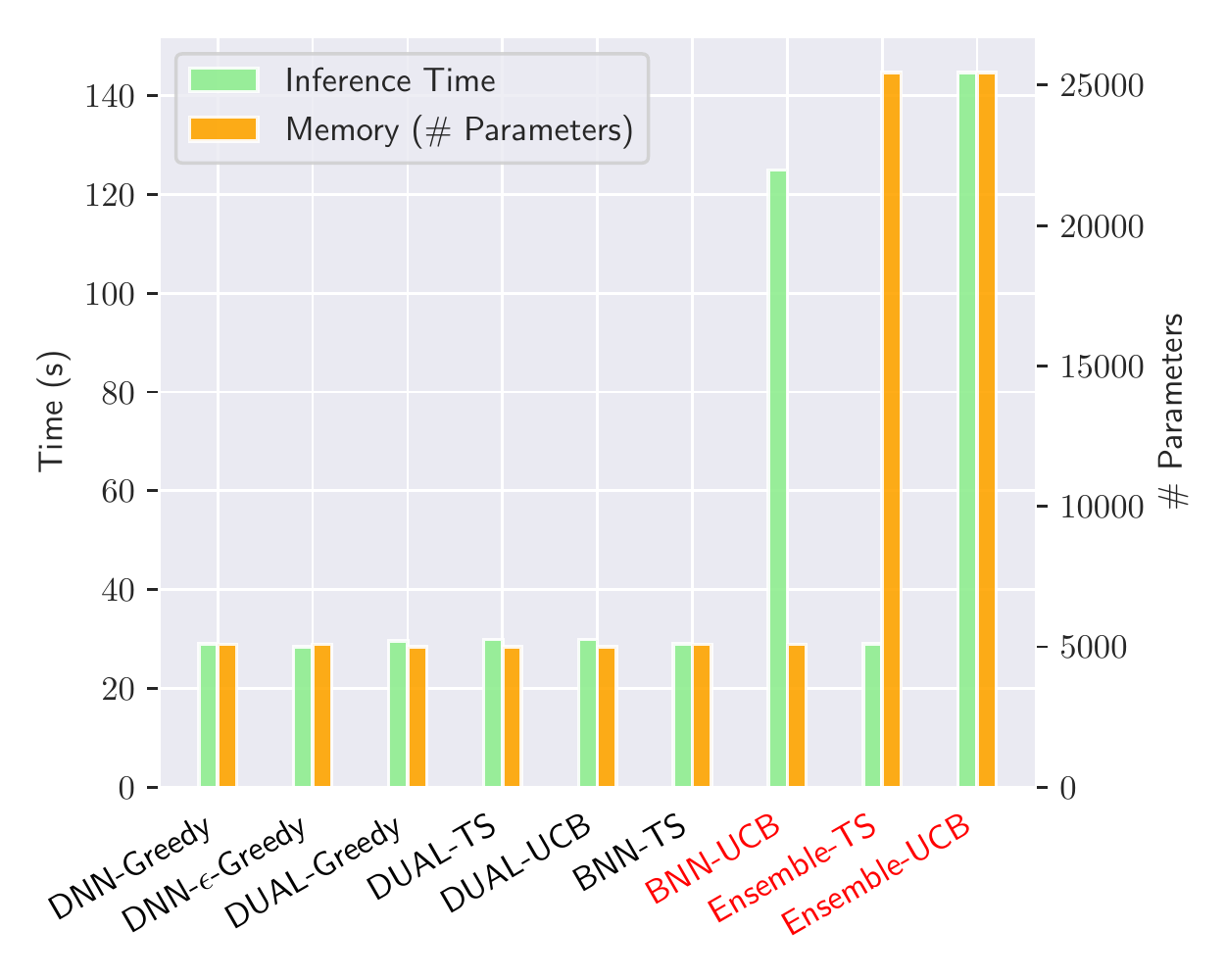} 
    \vspace{-.7cm}
    \caption{Inference time and memory costs}\label{fig:r6b-cost}
  \end{subfigure}
  \vspace{-.5cm}
  \caption{
    (a \& b) Cumulative social welfare (SW) and the daily CTRs obtained by different strategies on the Yahoo! R6B dataset. Results of each strategy are evaluated with 32 independent runs using the offline replay evaluation~\cite{li2011unbiased}.
    In (a) the red and black bars indicate the average and the median, respectively.
    (c) Inference time (of $80,000$ predictions) and memory cost of different strategies. 
    Strategies marked in red indicate that significant additional computational or storage overhead is required.
  }
  \label{fig:r6b}
  \Description{}
  \vspace{-.4cm}
\end{figure*}

\vspace{-.2cm}
\subsection{Long-Term Social Welfare Maximization}\label{sec:exp:explore}

We evaluate DUAL-based Ad-ranking strategies on the Yahoo! R6B dataset~\cite{yahoodata}, which contains around $28$ million lines of log data collected from the Today Module on Yahoo!'s frontpage within $15$ days of October 2011.
Each line consists of the following information:
\begin{itemize}[leftmargin=*]
  \item a $136$-dimensional multi-hot user feature;
  \item a candidate set of available news articles and the ID of the one displayed, which was randomly drawn from the candidate set;
  \item a binary label indicating whether the user clicked the article.
\end{itemize}
The dataset contains $652$ different articles and the average size of the candidate sets is $38$.
The random recommendations in the data collection process make the dataset perfectly unbiased, which allows us to evaluate the strategies with offline replay~\cite{li2011unbiased}.
We treat each news article as an Ad and assume that their bid prices are all equal to $1$ for simplicity.
In this case, the social welfare (SW) is just the number of total clicks.
We compare the following strategies:
\begin{itemize}[leftmargin=*]
  \item \textbf{DUAL-TS}, \textbf{DUAL-UCB}, and \textbf{DUAL-Greedy} as described in \secref{method:adrank}.
  For DUAL-UCB we choose $\kappa=1$ by manual search.
  \item \textbf{DNN-Greedy}.
  A deep CTR model is trained with observed samples under the purely supervised manner as described in \secref{back:ctr}.
  Ads are ranked according to its predictions, as described in \eqn{ecpm-greedy}.
  This approach corresponds to the common industrial practice.
  \item \textbf{DNN-$\epsilon$-Greedy}. A na\"ive approach to incorporate exploration into DNN-Greedy. With probability $1-\epsilon$, 
  it behaves the same way as DNN-Greedy; with probability $\epsilon$, it ranks Ads randomly.
  \item \textbf{BNN-TS} and \textbf{BNN-UCB}. 
  We learn a Bayesian deep CTR model using dropout approximation~\cite{gal2016dropout}, where dropout is performed in both training and inference time.
  The predictive distribution is approximated with multiple stochastic forward passes.
  The samples and the estimated mean/variance are then used in the TS and UCB strategies, respectively.
  \item \textbf{Ensemble-TS} and \textbf{Ensemble-UCB}.
  We train an ensemble of $5$ deep CTR models with bagging~\cite{murphy2012machine}.
  The predictions from a random picked model and the mean/variance of the predictions of all models are used in the TS and UCB strategies, respectively.
  \item \textbf{Random} selection of Ads. This acts as a trivial baseline.
\end{itemize}

For all deep structures, the same embedding technique is adopted to deal with the sparse inputs.
In specific, we use a $6$-dimensional embedding layer and sum pooling to convert the multi-hot user features to dense vectors and another $6$-dimensional embedding layer to encode the Ad IDs, resulting in $12$-dimensional compact embedding representations. 
For DUAL-based strategies, we use MLPs with $16$-$2$ FC layers after the embedding layers in the deep kernels and $M=8$ inducing points.
For all other methods, the deep CTR models are built with larger MLPs with $16$-$8$-$2$ FC layers after the embedding layers,
for fair comparison.


For each strategy,
we use the first $80,000$ log entries as the initial data to pre-train its CTR model(s).
After initialization, the remaining log entries are replayed to evaluate the strategies, according to the unbiased offline evaluation method proposed by \citet{li2011unbiased}.
Specifically, each strategy is asked to re-rank the candidate set in each log entry and to determine the winner Ad again.
If the winner Ad happens to be the one recorded as displayed in the log entry, the strategy will collect the corresponding Ad ID, the user feature, and the label as a new data sample, otherwise skip the log entry.
To simulate the daily/hourly updates of CTR models in real-world systems,
we update the strategies every $80,000$ log entries, resulting around $2,200$ data samples collected between each update and $337$ total updates during each replay. We choose the best hyperparameters for each strategy by manual search.

\figreft{r6b:swstd}{r6b:ctr} show the final cumulative SW and the daily CTRs obtained by each strategy. Results are averaged over $32$ repeated runs for each experiment.
First, we notice that DUAL-Greedy achieves better results than DNN-Greedy,
which verifies our conjecture in \secref{method:adrank}.
Then, we observe that DUAL-TS offers consistent performance improvements over greedy and $\epsilon$-greedy approaches, which suggests that the uncertainties provided by DUAL can help to balance the exploration-exploitation efficiently.
Specifically, DUAL-TS improves DNN-Greedy by $\mathbf{30.7\%}$ and improves DNN-$\epsilon$-Greedy by $\mathbf{10.0\%}$, benefiting from exploration.

Moreover, we find that DUAL-TS achieves better or comparable results compared to other TS/UCB strategies that use BNN or ensemble to estimate uncertainty.
As pointed out in \secref{relatedwork}, BNN and ensemble methods consume much more resources and are therefore impractical for large-scale real-world systems,
whereas DUAL does not incur additional computational and memory burden.
\figref{r6b-cost} shows the computational and memory costs of different strategies.
Notably, DUAL-TS yields a similar final cumulative SW result as Ensemble-TS with only $\mathbf{20\%}$ computational and memory overhead.

We also observe that DUAL-TS significantly outperforms DUAL-UCB.
Our conjecture is that the deterministic nature of DUAL-UCB might lead to ``greedy exploration'' that over explores a deterministic subset of Ads between successive updates, inhibiting the efficiency.
\vspace{-.2cm}
\subsection{Online A/B Testing}
We have deployed our method in the Alibaba display advertising platform and conducted an online A/B test for two weeks in October 2020.
Comparing with the previous state-of-the-art method used online, DUAL-TS increased the cumulative revenue by $\mathbf{8.0\%}$ and the social welfare by $\mathbf{8.2\%}$, during the A/B test.
To our best knowledge, this is the first large-scale industrial display advertising system that benefits from exploration-exploitation trade-off 
using uncertainty estimation in deep CTR models, without increasing computational overhead.
Remarkably, it had served the main traffic (over tens of billions of queries per day) during the Double 11 Shopping Festival 2020 in one of the most important scenes of Taobao,
affecting hundreds of millions of users.

It is worth mentioning that conducting online A/B testing for exploration problems is particularly difficult.
Methods in different experimental groups should be only allowed to learn from the samples produced by their own models and strategies, in order to prevent leakage of information gained during exploration.
To this end, we developed a \emph{sample-isolation mechanism} in the Alibaba display advertising system, which helps label the samples with different models and strategies in real-time.
Such a mechanism allows us to fairly evaluate and compare the experimental results.

\vspace{-.2cm}
\section{Conclusions and Discussions}

In this paper, we focus on the exploration problem in advertising systems.
We propose DUAL for learning deep CTR prediction models,
which enables efficient estimates of predictive uncertainties while maintaining the advantages of deep model structures.
We then propose
DUAL-based Ad-ranking strategies that balance the exploration-exploitation trade-off in advertising systems,
leading to optimized long-term utilities such as the social welfare.
Experimental results on public datasets demonstrate the effectiveness of the proposed methods. An online A/B test deployed in the Alibaba display advertising platform shows remarkable improvements in both revenue and social welfare.

We hope this work will help motivate researchers and industrial practitioners to pay more attention to the exploration problem in advertising systems and to explicitly consider how the data samples are generated when training CTR models,
for which our work is just a preliminary attempt.
Our future work includes:

\noindent
\textbf{Extensions of DUAL.}
We like to investigate better ways to express suitable inductive biases in user behaviors with kernels, such as graph kernels~\cite{li2013recommendation}, to promote the capability of DUAL.
Designing better parameterization of the variational distribution to improve flexibility and scalability~\cite{shi2020sparse} is another interesting direction.

\noindent
\textbf{Extensions of DUAL-based Strategies.}
Recent progress~\cite{zhu2017optimized,wu2018budget,zhao2018deep} has also studied more complex strategies under constraints such as the Return on Investment (ROI) and the advertising budgets.
Combining DUAL with these strategies to facilitate long-term benefits under constraints is an interesting challenge.

\noindent
\textbf{Exploration in Retrieval.}
To alleviate the computational burden in large-scale advertising systems, a fast retrieval model~\cite{zhuo2020learning,zhu2018learning} is often needed before the fine-grained CTR prediction, which can also suffer from lack of exploration when learned from user feedback. It is interesting and valuable to adapt DUAL to learn large-scale retrieval models, which can potentially enable exploration in the retrieval stage and amplify the benefits of exploration.

\begin{acks}
  We would like to thank Chongxuan Li, Jiaxin Shi, and Kun Xu for their helpful comments and discussions.
\end{acks}
\bibliographystyle{ACM-Reference-Format}
\bibliography{ref}


\end{document}